\documentstyle[preprint,aps,eqsecnum]{revtex}
\begin{document}
\vspace{2cm}
\title
{Schwinger Particle-Production Mechanism for a
Finite-Length Flux Tube with Transverse Confinement}
\author{
Cheuk-Yin Wong }
\address
{Oak Ridge National Laboratory, Oak Ridge, TN 37831}
\author{
Ren-Chuan Wang }
\address
{University of Science and Technology of China, Hefei, China}
\author{
Jian-Shi Wu  }
\address
{Fayetteville State University, Fayetteville, NC 28301}
\date{\today}
\maketitle
\vspace{2cm}
\begin{abstract}
Previous results for the pair production probability in a strong
electric field with a finite longitudinal separation are generalized
to the case of a finite-length flux tube with transverse confinement.
The threshold length of the flux tube, below which pair production
cannot occur, increases as a result of transverse confinement.
\end{abstract}

\pacs{ PACS number: 12.90.+b  }
\newpage

\narrowtext

\section{INTRODUCTION}
\label{sec:intro}

The Schwinger particle production mechanism has found  applications in many
branches of physics  \cite{sch51}-\cite{gle83}.  It was originally derived
\cite{sch51} using the Green's function and gauge invariance for the case
where the strength of the Abelian electric field is a  constant in the whole
configuration space. For many physics applications, the field strength exists
only in a limited region of space.   For example,  in  quantum chromodynamics
(QCD), the color flux tube is a useful concept, as it explains many qualitative
features of nonperturbative phenomena such as confinement, Reggeon
trajectories, and particle production \cite{cas79}-\cite{gle83}. The
potential energy of a flux tube
is linearly proportional to
their separation.  Hence, we can approximate the field between the
particle-antiparticle pair by a constant  Abelian electric field, as in the
case studied by Schwinger.
The electric field strength is non-zero only in a
finite spatial region. It is of interest   to consider the case of the
Schwinger production mechanism for which the electric field is restricted to
a finite longitudinal and transverse region.

Previously, Wang and Wong (\cite{wan88}, which we shall call I) studied the
Schwinger particle production mechanism for the case  where an electric field
exists only between two parallel condenser plates at a finite longitudinal
separation. The transverse dimension is taken to be infinite.
The Klein-Gordon and the Dirac equations are solved  in
this field. The transmission amplitude, which is related to the probability for
pair production, was worked out and written explicitly in terms of parabolic
cylinder functions for bosons and confluent hypergeometric functions for
fermions \cite{wan88}. It is found that pair-production occurs only when the
longitudinal separation is greater than
$m_0/\sigma$ where $m_0$ is the rest mass of the produced particle and
$\sigma$ is the string tension. The
pair-production probability is much suppressed when the separation just exceeds
the threshold value.  The probability also shows oscillations due to the
finite separation.
The same problem was examined by Martin and Vautherin using Schwinger's
Green's  function formalism and  Balian and Bloch's multiple reflection
expansion  method \cite{mar88}. Martin and Vautherin also studied the
effect for the case with a transverse boundary.

Subsequently, Pavel and Brink \cite{pav91}  introduced the scalar potential
$m(r)$ in the transverse direction to  study a flux tube of finite transverse
radius.   They examined a flux tube with a sharp transverse
boundary and infinite
longitudinal length,  which was also studied by Sailer $et~al.$ \cite{sai90}
and Sch\"ofeld $et~al.$ \cite{sch90}. They obtained pair-production probability
which was found to be suppressed due to the finite radius of the tube.   Sailer
$et~al.$ \cite{sai90}  and Sch\"ofeld $et~al.$ \cite{sch90} also calculated the
transverse momentum distribution for a flux tube with a sharp boundary. In a
recent work, Gatoff and Wong  \cite{gat92} studied the low $p_{{}_T}$ spectra
in terms of a smooth-varying transverse potential.
The Dirac equation was
solved by using the method of separation of variables of Pavel and Brink
\cite{pav91}.      The effect of including an additional
scalar potential  rising linearly with the absolute magnitude of the
longitudinal coordinate was investigated by
Warke and Bhalerao \cite{war92a,war92b}.  Pair production in the color
dielectric model with a smooth transverse boundary has been
investigated by Flintoft and Birse \cite{fli92}.

Previous works of Schwinger mechanism with transverse confinement deals
with a color flux tube of infinite longitudinal length.
We wish to investigate in this paper
the case of the Schwinger particle production mechanism for a
finite-length
flux tube with transverse confinement for both bosons and fermions.
As the pair production probability depends on the transmission
amplitude for  longitudinal incident waves, it is necessary to introduce the
incident, the reflected and the transmitted waves.  In Refs. \cite{pav91} and
\cite {gat92}, no explicit prescription was given to separate the longitudinal
incident waves and the transmitted waves.
In the work of Pavel and Brink \cite{pav91} for fermions, the
longitudinal component of the wave function are different for
different spinor components (see Eq.\ (3.7) and (3.11) of Ref.\
\cite{pav91}) and the formulation of the transmission and reflection
of incident waves to obtain the pair production probability was not
carried out.  We wish to report here a different method of
separating variables, by following the representation of our
earlier work in Ref. \cite{wan88}. With this representation,
the longitudinal wave function can be completely factorized out
from the transverse wave function in the spinor form (see Eqs.
({\ref{eq:psiphi}) and
({\ref{eq:psi2}) below). Therefore, we can introduce
explicitly the longitudinal incident, reflected
and transmitted waves.   This method provides a formalism for calculating the
transmission amplitude for the general case of a flux tube of finite
longitudinal length $L$ and a smooth varying transverse confining scalar
potential.   The separation of the incident and transmitted waves turns out to
be just a generalization of the previous results in I (Ref.
\cite{wan88}). The results on the transmission amplitude  and reflection
amplitude from Ref. \cite{wan88}  can be used directly for the case with
transverse confinement by replacing the continuous transverse mass
$m_{\perp}$ of I with the
discrete eigenvalues when there is a transverse confining potential.
Thus, in the  application of the Schwinger
mechanism for particle production, the proper mass involved in the
production process is not the rest mass but the transverse mass,
including the effect of the transverse zero-point motion.

\section{Bosons in a Strong Field }

We consider first a boson of mass $m_0$ in a linear vector potential
$A=(A_0(z),0,0,0)$ where $A_0$ depends only on $z$ and is of the form
\begin{eqnarray}
\label{eq:pot}
A_0(z)=\cases{0\;\; & for $z\le 0\qquad\qquad$  ~~(region I) \cr
-\kappa z  & for $0\le z\le L\;\;\;$  ~~~~~(region II) \cr
-\kappa L \;\; & for $\;\, L \le z\qquad\qquad$  (region III)~. \cr}
\end{eqnarray}
Such a linear potentail arises, for example,
in an approximate description of
pair production in a flux tube of length $L$ in QCD, where
the strength parameter
$\kappa$ is related to the string tension $\sigma$ ($\kappa=2\sigma$  in
\cite{gle83,wan88,won94}).
We study  the case where the transverse motion of the boson is
restricted and described by a transverse scalar
potential  $m(r)$ which  includes the rest
mass $m_0$.
The equation of motion of the boson is the
Klein-Gordon equation:
\begin{eqnarray}
[(p-A)^2-m^2(r)]\phi(r,\varphi,z)=0,
\end{eqnarray}
The solution of the equation can be written in the form
$
\phi(r,\varphi,z)=R_\nu(r)e^{i\nu\varphi}f(z),
$
which allows the equation to be separated into the set
 of coupled equations
\begin{eqnarray}
\label{eq:bos1}
[ -\partial_z^2+{m_{\perp}}^2 -(p_0- A_0)^2 - i \partial_z A_0(z)] f(z)=0,
\end{eqnarray}
\begin{eqnarray}
\label{eq:bos2}
\biggl \lbrack
- {1\over r} {\partial \over \partial r} r {\partial \over \partial r}
+{\nu^2 \over r^2} +m^2(r) -{m_{\perp}}^2 \biggr \rbrack R_{\nu}(r)
=0 \,,
\end{eqnarray}
where $m_{\perp}$ is the constant of separation and
plays the role of the transverse mass.
For a given transverse
quantum number $\nu$, one solves for the transverse wave function
$R_\nu(r)$ in Eq. (\ref{eq:bos2}) with the eigenvalue $m_\perp^2$,
subject to the
boundary condition that the wave function $R_{\nu}(r) \rightarrow 0$ as
$r\rightarrow \infty$.
For the case of a sharp boundary with
a radius $r_0$,
the minimum eigenvalue of $m_\perp$
has the
value
$
m_{\perp}=\sqrt{m_0^2 + (2.404/r_0)^2},
$
which increases as the size of the flux tube $r_0$ decreases.
In another example, with
$m^2(r)=m_0^2 + \sigma^2 r^2$,
the lowest value of $m_{\perp}$ is
$
m_{\perp} = \sqrt{m_0^2+2 \sigma},$
which contains the transverse zero-point energy.
In the special case when the boson is a gluon with a zero rest mass,
the gluon  acquires an effective mass $m_\perp$ as a result of
transverse confinement.
Finally,
for two parallel plates
for which $r_0\rightarrow \infty$, there is no
restriction in the transverse motion. The solution of $R(r)$  contains
plane waves
in the transverse directions and
$m_\perp^2=m_0^2+p_{{}T}^2$.

After obtaining the eigenvalue $m_{\perp}$ from the transverse motion,
the dynamics of the boson  for the longitudinal motion is determined
by Eq. (\ref{eq:bos1}).
One notes that Eq. (\ref{eq:bos1}) is
the same as the boson case in I by replacing the continuous
transverse mass $m_\perp$
in I with discrete $m_{\perp}$.
Because of this simple relationship, one can carry out the same steps  of
derivation so that previous results in I for the transmission  and the
reflection of an incident wave coming from the right in Region I  for two
parallel condenser plates can be used directly here.

\section{
Fermions in a Strong Field }

To examine the production  of fermion-antifermion pairs
in a flux tube
we study the motion of a fermion in the Abelian
gauge field $A_\mu=(A_0(z),0,0,0)$, and the transverse confining potential
$m(r)$, as introduced in Section II.
The Dirac equation for the fermion in cylindrical coordinates $(r,\varphi,z)$
is:
\begin{eqnarray}
\label{eq:dir}
\biggl \lbrack \gamma \cdot \Pi -m(r)
\biggr \rbrack \psi(r,\varphi,z)=0,
\end{eqnarray}
where
$\gamma\cdot \Pi=\gamma^\mu \Pi_\mu$
and $\Pi_\mu=p_\mu- A_\mu.$
A fermion and an antifermion
are spontaneously produced in this field when a
fermion in a negative-energy state
tunnels from the negative energy continuum to the positive energy
continuum \cite{wan88}.
Generalizing the results of Ref. \cite{wan88},
we seek a solution of the Dirac equation
(\ref{eq:dir}) in the form
\begin{eqnarray}
\label{eq:psiphi}
\psi(r,\varphi,z)=
\biggl [\gamma \cdot \Pi +m(r)\biggr ]\phi(r,\varphi,z).
\end{eqnarray}
The equation for $\phi$ is
\begin{eqnarray}
\label{eq:ephi}
\biggl \lbrace (p_0-A_0)^2- \bbox{p}^2-m^2(r)
+i\alpha^3 \partial_z A_0(z)
+i \biggl [(\gamma^1\partial_1+\gamma^2 \partial_2)m(r)\biggr ]
\biggr \rbrace  \phi(r,\varphi,z) = 0.
\end{eqnarray}
We note that
$\biggl [\alpha^3, J_z\biggr ] = 0,$
where
 $J_z = -i \partial / \partial\phi + \sigma_z/2 $
 is the third component of the angular momentum operator.
Furthermore, both $J_z$ and $\alpha^3$ commute with the operator
acting on $\phi(r,\phi,z)$ in Eq.\ ({\ref{eq:ephi}).
Therefore, we may choose the $\phi(r,\varphi,z)$ to be
factorized as
\begin{eqnarray}
\label{eq:psi2}
\phi_{\eta \nu}(r,\varphi,z)=
f_{\eta\nu}(z) \rho_{\eta\nu}(r,\varphi),
\end{eqnarray}
with
$\rho_{\eta\nu}(r,\varphi)$ to be
simultaneous eigenfunctions of $\alpha^3$ and $J_z$.
Upon using the representation in Ref. \cite{wan88},
the eigenfunction of $\alpha^3$
satisfying
$\alpha^3 \mu_\lambda=\eta_\lambda \mu_\lambda$ are
\begin{eqnarray}
\label{eq:rho}
  \mu_1  = {1 \over \sqrt{2}}
             \pmatrix{ 1       \cr
                       0       \cr
                       1       \cr
                       0       \cr},
{}~~\mu_2  = {1 \over \sqrt{2}}
             \pmatrix{ 0       \cr
                       1       \cr
                       0       \cr
                      -1       \cr},
{}~~\mu_3  = {1 \over \sqrt{2}}
             \pmatrix{ 1       \cr
                       0       \cr
                      -1       \cr
                       0       \cr},
{}~~\mu_4  = {1 \over \sqrt{2}}
             \pmatrix{ 0       \cr
                       1       \cr
                       0       \cr
                       1       \cr},
\end{eqnarray}
with $\eta_{1,2}=+1$ and
with $\eta_{3,4}=-1$.
Thus, the eigenfunctions of $J_z$ satisfying $J_z
\rho_{\eta \nu}=(\nu+\sigma_z/2) \rho_{\eta \nu}$ are
\begin{mathletters}
\begin{eqnarray}
\rho_{\eta \nu} = g_{1\nu}(r) e^{i\nu\phi} \mu_1
              -g_{2\nu}(r) e^{i(\nu+1)\phi} \mu_2 ~~~{\rm for}~~\eta=+1,
\end{eqnarray}
\begin{eqnarray}
\rho_{\eta \nu} = g_{1\nu}(r) e^{i\nu\phi} \mu_3
              +g_{2\nu}(r) e^{i(\nu+1)\phi} \mu_4 ~~~{\rm for}~~\eta=-1.
\end{eqnarray}
\end{mathletters}

The Dirac  equation can be easily separated into the set of
equations
\begin{eqnarray}
\label{eq:ez1}
[-\partial_z^2 +{m_{\perp}}^2- (p_0- A_0)^2  +\eta  i
\partial_z A_0(z)] f_{\eta\nu}(z)=0,
\end{eqnarray}
\begin{mathletters}
\begin{eqnarray}
\label{eq:tr1}
\biggl [\bbox{p}_\perp^2(\nu) + m^2(r) - {m_{\perp}}^2 \biggr ]g_1(r)
       = i{{\partial m(r)}\over{\partial r}} g_2(r) ,
\end{eqnarray}
\begin{eqnarray}
\label{eq:tr2}
\biggl [\bbox{p}_\perp^2(\nu+1) + m^2(r) - {m_{\perp}}^2\biggr ]g_2(r)
       =- i{{\partial m(r)}\over{\partial r}} g_1(r) ,
\end{eqnarray}
\end{mathletters}
where
\begin{eqnarray}
\bbox{p}_\perp^2(\nu)
= -{{1}\over{r}}
{{\partial}\over{\partial r}}(r{{\partial}\over{\partial r}})
                        + {{\nu^2}\over{r^2}} \,.
\end{eqnarray}
When
 $ m(r)=\rm{constant}, $ the two functions are decoupled.

Note that in the transverse direction the positive and negative energy
states are degenerate for $\rho_{\eta\nu}(r,\varphi)$, and
$\bigl [\eta m_{\perp} \beta - \bbox{\gamma}_\perp \cdot \bbox{p}_\perp + m(r)
\bigr ]\rho_{\eta\nu}(r,\varphi)$
would automatically produce two different solutions with $\eta=\pm 1$
for the transverse
Dirac equation, because of the sign difference in the $m_{\perp}$ term.
The advantage of separating the variables in the form of Eqs.\
(\ref{eq:psiphi}) and (\ref{eq:psi2}) is that as $f_{\eta \nu}$ is
factorized out, the separation of the corresponding components of the
wave function in terms of the transmission and reflection amplitudes
can be easily carried out.
To determine  these amplitudes,
we match the wave function $f_{\eta\nu}(z) $ and
$f'_{\eta \nu}(z)$ at $z = 0$ and $z = L$.
It  can be shown after tedious algebra that the
transmission probabilities for $f_{\eta=+1,\nu}(z)$ and $f_{\eta=-1,\nu}(z)$
states
are equal. Therefore, it is necessary to discuss only the case of
$f_{\eta=+1,\nu}(z)$ explicitly.

Since the transformation (\ref{eq:psiphi}) from $\phi$ to $\psi$ is through
an operator
$[(p_0 - A_0(z))\beta - \bbox{\gamma} \cdot \bbox{p}+ m(r)]$, which contains
the operator
``$\partial / \partial z$", the continuity of the wave function
$\psi$ at $z=0$ and $z = L$ will require continuity for
both $f_{\eta\nu}(z)$ and $f_{\eta\nu}'(z)$ at these boundaries.

Making the transformation from $z$ to $\xi$ by
$\xi=(\kappa z +E)\sqrt{2/\kappa}$,
the longitudinal
solutions in Regions I and III are just harmonic waves.  For Region II, the
longitudinal equation (\ref{eq:ez1}) becomes
\begin{eqnarray}
\label{eq:fer}
\biggl \{ -{d^2 \over d\xi^2} + (a +\eta  {i \over 2} )
- {1\over 4} \xi^2
\biggr \} f_{\eta\nu} (\xi) =0,
\end{eqnarray}
with
\begin{eqnarray}
\label{eq:fer2}
a={ {m_{\perp}}^2 \over 2\kappa}.
\end{eqnarray}
Eqs.\ (\ref{eq:fer}) and ({\ref{eq:fer2}) are identical to Eqs.\ (3.9)
and (2.8) of I.   Hence, the solutions in I can be used directly here
by making the
replacement
\begin{eqnarray}
m^2+p_x^2+p_y^2 ~~({\rm of~Ref.~3}) ~~\rightarrow ~~{m_\perp}^2,
\end{eqnarray}
as shown by Pavel and Brink \cite{pav91}.

\section{The Pair-Production Probability }
A particle-antiparticle pair is
spontaneously produced in the strong field when a
particle in a negative-energy state $E$ with a momentum $-|p_z|$
tunnels from the negative energy continuum to the positive
energy one \cite{wan88}.
By choosing the incident amplitude $I$ to be unity,
the transmission amplitude $T(E)$ is the probability amplitude for the
transmission of a particle from Region I to  Region III
with the accompanying creation of an antiparticle in Region I.
For an incident particle with an energy $E$ the probability for a pair
production is
then $|T(E)|^2$.
The threshold flux tube length, below which no pair production
can take place, is $2 m_\perp/\kappa$.  Because $m_\perp$ is greater
than the rest mass $m_0$ due to the zero-point motion,  the threshold
length of the flux tube increases as a result of transverse confinement.
For a
given length
$L$,  we need  to consider only a
 finite number of transverse excited states
with $m_\perp \le {1\over 2} k L$.
The longer the flux tube, the greater will be the
number of  transverse states  that can  be produced.

In the case without a
boundary, the transverse
mass $m_\perp^2= m_0^2+{\bbox{p}}_\perp^2$
is a continous variable, but
 in a flux tube the transverse
mass assumes discrete values.  Therefore, in evaluating
the pair production rate
the phase-space integral in the transverse direction,
    $$A \sum_{spin} \int {{d^2p_\perp}\over{(2\pi)^2}},$$
with the transverse area $A$,
is replaced by a summation over
the finite number of transverse states,
$\sum_{\eta \nu}.$ For each transverse state, the
integral in the longitudinal momentum is restricted by
$E=-\sqrt{p_z^2+{m_{\perp}}^2} \ge m_\perp-kL$,
or $$ 0 \le | p_z | \le \sqrt{kL(kL-2 m_\perp)}.$$

With the replacement of the continuous integral in $\bbox{p}_\perp$
by the discrete summation
 in  Eq. (4.12) in \cite{wan88},
the particle production rate becomes \cite{ft}

$$ {{\Delta N}\over{\Delta t}} =\sum_{\eta\nu}
\int r dr d\varphi
\int_0^{\sqrt{kL(kL-2m_\perp(\nu))}}  {{p_zdp_z}\over{2\pi E}}
\psi^\dagger_{\eta\nu}(r,\varphi,z) \psi_{\eta\nu}(r,\varphi,z)
{ {{v_R}\over{v_L}}} |T^{(\eta)}_\nu(E)|^2,  $$
where
$$\psi_{\eta\nu} (r,\varphi) =
[ E\beta - \bbox{\gamma} \cdot \bbox{p} + m(r) ]
\phi_{\eta \nu}(r,\varphi) a_I e^{-i k_L z}.$$
Since $\phi_{\eta\nu}(r,\varphi)$ is properly normalized, and
$\psi^\dagger_{\eta\nu}(r,\varphi,z) \psi_{\eta\nu}(r,\varphi,z)$
is independent of $z$, we have
$$\int r dr d\varphi
\psi^\dagger_{\eta\nu}(r,\varphi,z) \psi_{\eta\nu}(r,\varphi,z)
= \int r dr d\varphi \phi_{\eta\nu}^\dagger(r,\varphi)
\phi_{\eta\nu}(r,\varphi) = 1 .$$
Therefore, we have
$$ {{\Delta N}\over{\Delta t}}
= \sum_{\eta\nu} \int_0^{\sqrt{kL(kL-2m_\perp(\nu))}}
 {{p_zdp_z}\over{2\pi E}}
{ {{v_R}\over{v_L}} } |T^{(\eta)}_\nu(E)|^2.$$
These limits are
important for a numerical calculation, especially for the rate
near the threshold or for high transverse excitations.
In the continuous limit where the transverse area $A$ is large,
the formula for the rate of pair production given in
Ref. \cite{wan88}
is recovered except for the explicit limits on
the integral of $p_z$.

\section{ Conclusions and Discussions}

It was recognized by Pavel and Brink for an infinitely
long flux tube and fermion pairs  that the pair production probability
for the case with transverse confinement can be obtained by replacing
the continuous mass with a discrete set of $m_\perp$ \cite{pav91}.
Generalizing this result to the case with a finite longitudinal length,
for both fermion and boson pairs, the pair production probability are
just given
by previous
results obtained in I,  with the replacement of
the continuous
transverse mass $m_\perp$  by a set of discrete $m_{\perp}$  obtained
in  an eigenvalue equation in the transverse degrees of freedom.
A different transverse excitation is represented by
a different
mass $m_{\perp}$ which
contains contributions from the rest mass (current mass) and the additional
zero-point energy arising from  tranverse
confinement.
Pair production  occurs only when the longitudinal separation is
greater than the threshold length $2m_{\perp}/\kappa$ which increases
as a result of transverse confinement.
Furthermore, the pair-production probability is much suppressed when
the separation just exceeds the threshold value.  The probability also
shows oscillations due to the finite separation \cite{wan88}.
In contrast,
in the case without transverse confinement,
 the mass which needs to
tunnel through the barrier in the longitudinal direction
$m_\perp=\sqrt{m_0^2 + p_T^2}$ is a continuous
variable.  The lowest transverse mass with $p_T=0$
is just the rest
mass of the particle $m_0$.

The above results shed light on the values of masses one should use in
the application of
the Schwinger mechanism to examine $q \bar q$ production \cite{cas79,won94}.
There, it is often a question
whether one should use the current quark mass (of a few MeV for $u$ and $d$
quarks)
or the constituent quark mass (which is of the order of a few hundred MeV),
for the threshold of pair production. Our analysis of the
transverse boundary
indicates that the proper mass one should consider for the Schwinger
mechanism is neither the current mass nor the constituent mass but
the transverse mass $m_{\perp}$ which depends on the current quark
mass $m_0$,
the  finite
transverse dimension of the flux
tube, and the transverse state quantum number $\nu$ of the particle.
It  includes the zero-point energy of
the transverse motion, which is of the order of $\hbar/$(flux tube
radius).
With a flux tube radius of the  about 0.5 fm, the transverse zero-point
energy is of the order of a few hundred MeV.  The transverse mass
$m_{\perp}$
is therefore
about a few hundred MeV,
for the production of particles in
the lowest transverse state.

\acknowledgments
The authors would like to thank Dr. G. Gatoff for helpful discussions.
This research was supported in part by the Division of Nuclear  Physics, U.S.
Department of Energy under Contract No. DE-AC05-84OR21400  managed by  Martin
Marietta Energy Systems, Inc., and in part under Contract No.
DE-FG05-94ER40883.


\begin{references}
\bibitem{ft}
In Ref. \protect\cite{wan88}, Eq. (4.11) has a misprint;  the expression
``$|{\bf J}_{{}_T} |/|{\bf J}_{{}_L}|$"
on  the left hand side should read
``$|{\bf J}_{{}_T} |/|{\bf J}_{{}_I}|$", where
the subscript $I$ stands for the incident  current.
\end{references}
\end{document}